\documentstyle[12pt,aasms4]{article}
\def \yskip{\penalty-50\vskip3pt plus 3pt minus 2pt}
\def \pp{\par \yskip \noindent \hangindent .4in \hangafter 1}
\def \abc#1#2#3#4 {\pp#1, {\sl#2}, {\bf#3}, #4}
\def \blank {\lower 5pt\hbox to 0.75in{\hrulefill}}
\def \kms{km~$\rm{s}^{-1}$}
\def \cc{$\rm{cm}^{-3}$}
\def \lam{$\lambda$}

\def \Hb{H$\beta$}

\def \msolar{M$_{\odot}$}

\newfont{\rten}{cmr10}

\begin{document}
\parskip 10pt
\baselineskip=20pt

\title{The Optical Spectroscopic Evolution of V1974~Cygni (Nova Cygni 1992)}
\vspace*{2.5cm}
    
\author{Amaya Moro-Mart\'{\i}n}
\affil{Steward Observatory, 933N. Cherry Ave., Tucson, AZ 85721}
\affil{email: amaya@as.arizona.edu}
     
\author{Peter M. Garnavich}
\affil{University of Notre Dame, Nieuwland Science Hall 213, Indiana, 46556}
\affil{email: Peter.M.Garnavich.1@nd.edu}

\affil{\&}      

\author{Alberto Noriega-Crespo}
\affil{SIRTF Science Center, IPAC, CalTech 100-22, Pasadena, CA 91125}
\affil{email: alberto@ipac.caltech.edu}
 
\begin{abstract}

Optical observations of nova V1974 Cygni (Cygni 1992), spanning 
a 4 year period, have been used to study its spectroscopic evolution.   
The data cover a wavelength range from $\sim 3200 - 8000$ \AA~ 
and follows the nebular evolutionary phase of the ejecta.
  
We have modeled the integrated fluxes by means of the photoionization code 
CLOUDY. The models were run at a {\it fixed} abundance value for the
most prominent elements (i.e. H, He, C, O, N, Ne, Fe, etc) over the entire
time sequence. It is possible to constrain from this simple model some of 
the physical conditions of the gaseous emitting region, like temperature
and density.

Compared with previous studies of the gas abundances of the heavy elements, 
we found that smaller enhancements of S, N and Ar, and comparable values 
for O and Fe, are able to reproduce 
the observations.

The time evolution of the surface temperature of the ionizing source and
the high-ionization iron lines [Fe VII] 6087 \AA~ and [Fe X] 6374 \AA, 
is similar to what it is observed in the soft X-rays.

The early line profiles can be reproduced using a simple kinematical model 
consisting of an equatorial ring and polar caps, expanding at
a velocity of $\sim 1100$ ~km/s. This simple model also approximates
the structure of the resolved shell observed by HST.

Considering the complicated structure of the shell, the lack of well
defined values of its gas density and our limited knowledge of the time
evolution of the surface temperature of the photoionization source,
the comparison between models and observations agrees remarkably well.

\end{abstract}


\section{Introduction}

Nova outbursts are associated with close binary systems
in which the secondary star fills its Roche lobe and transfers gas
onto the companion white dwarf via an accretion disk 
(\markcite{geh88}Gehrz, 1988).  
This hydrogen rich
gas accumulates onto the surface of the white dwarf until temperatures and 
densities are high enough to ignite thermonuclear reactions.  A thermonuclear
runaway process takes place leading to the ejection of gas at hundreds or
thousands of kilometers per second (\markcite{sai96}Saizar et al. 1996). 
This expanding gas has a chemical 
composition which is more metal rich than Solar, and later on is photoionized 
by the hot underlying white dwarf.
The thermonuclear runaway model can account for some
CNO enhancement abundances observed in novae. Some other heavy element 
abundances, however, as the ones measured in Nova Cygni~1992 
(\markcite{aus96}Austin et al. 1996; hereafter A96), can apparently only be 
explained by a mixing of the accreted envelope with white dwarf's core 
material (\markcite{sta86}Starrfield et al. 1986 and 
\markcite{tru86}Truran and Livio 1986).

Two types of nova are commonly recognized: CO novae, in which the progenitor
is a carbon-oxygen white dwarf, and ONeMg novae, involving an oxygen-neon-
magnesium white dwarf. Novae Cygni 1992 belongs to this second type since the 
most prominent emission lines are [Ne~III] \lam\lam3869/3967, [Ne~V] 
\lam\lam3346/3426~ and [O~III] \lam\lam4959/5007 
(\markcite{mat93}Matheson et al. 1993).

Here we use spectroscopic observations taken over a 4 year 
period, coupled with photoionization models, in order to study the time 
evolution of the nebular line
spectra of Nova Cygni 1992 from day 85 to day 1581 after the outburst. 
The models provide
the white dwarf's temperature and the hydrogen density and chemical
composition of the emitting shell under the simple approximations
of uniform and spherically symmetric shell. The spectral line structure
and HST images show a more complex morphology (\markcite{par95}Paresce et al. 1995),
which suggests one must interpret these results with caution.

\section{The Observations}

The data were obtained primarily with the
Dominion Astrophysical Observatory 1.8-m Plaskett telescope, but as the nova faded
the Kitt Peak National Observatory 2.1-m and the Multi-Mirror Telescope were
employed.
The log of the observations is shown in Table 1. The nebular observations began on May 15, 1992
(85 days after the outburst) and continued until June 1996 (1581 days after maximum).  
The data are all long-slit spectra using a CCD as a detector. The spectral
images were reduced using standard IRAF packages to remove amplifer
bias and flat field variations as well as to extract and calibrate the spectra. Moderate 
resolution combined with good 
wavelength coverage was achieved by taking spectra of the nova and standard stars at a variety 
of grating tilts, taking care that consecutive spectra overlapped in wavelength.
The spectra were then reduced individually and combined by scaling to match the overlap sections.  

For the DAO observations, the slit width was 1.5$''$ to 2$''$ and standard
stars were observed near the same airmass as the nova, which reduces the
effects of slit losses caused by differential refraction. The KPNO
and MMT spectra were taken at the parallactic angle. The major source
of error in the observed line ratios comes from combining spectra
taken with different grating tilts as this error can accumulate over
a large span of wavelengths. We estimate that ratios of lines with
small wavelength separations ($<$100 nm) are good to 10\% while
lines ratios spanning $>$300 nm are accurate to about 30\% .

\section{Line Shapes}

The nebular emission lines of V1974~Cyg were broad ($\sim 2000$ km/s
FWHM) with a symetric shape that persisted for the first two
years after discovery.  The evolution of the spectral line shape is best exemplified by 
[O~III] \lam\lam4959/5007 emission.  Figure~1 displays the [O~III] line in 1992.
Early on the lines were extremly broad, suggesting a fast wind dominated
emission and smeared out the kinematic features.  Through out 1992, the [O~III] line
narrows and the individual features become more clearly defined.  Figure~2 shows
that between 1992 and 1995 the [O~III] lines become even more narrow, and
the symmetry dissolves into many knots as the emission becomes dominated
by low density gas that is slow to recombine.

The striking symmetry in the early spectral features suggests a simple 
geometrical structure of the ejecta and implies that
the line profile can be modeled in a straightforward
way by following the prescription developed by \markcite{hut72}Hutchings (1972).
The ejecta from novae in Hutchings' model consists of an expanding
ring and polar caps, which inherently reflect the binary symmetry of the 
system. Our basic kinematical model consists of a shell of emission expanding at
a velocity of 1100$\pm 100$\kms. The emission is enhanced in the equatorial region and polar
regions and the viewing angle is 35$^\circ$ from the pole. Figure~3 shows a
comparison between a simple kinematical model and  the observations of the
[O~III] \lam5007 line profile for three epochs in 1992. This single-velocity
equatorial ring/polar blob model reproduces the basic spectral features, except
for the lowest velocity emission peak and the broad wings of the earliest data.
 
Figure~4 displays a projection onto the sky at the [O~III]  
\lam\lam5007 emission compared with an archival 1994 
HST FOC observation of Cygni 1992 at the same wavelength range.
The model image perhaps exaggerates the polar caps, but overall it resembles 
the observations. The parameters of the model, the same for the spectrum and 
image, are summarized in Table 2. The line profiles have also been modeled by 
\markcite{aus95}Austin (1995) and by \markcite{hay96}Hayward et al. (1996) 
(hereafter H96), based on a more complete physical model. 
Note that hydrodynamical simulations of nova shells
by \markcite{gill99}Gill \& O'Brien (1999) fail to reproduce
the equatorial and polar features that appear to be common
in nova shells.

\section{The Photoionization Models}

The emission line fluxes were modeled using the photoionization code CLOUDY,
version 84.12a
(\markcite{fer93}Ferland 1993). 
The code simultaneously solves the equations of thermal and statistical 
equilibrium for a spherical shell of a dilute gas which is being photoionized by an 
underlying thermal source. The input parameters for the code 
are the temperature and luminosity of the source (assumed to be a blackbody) 
and  those describing the characteristics of the shell, i.e. the hydrogen 
density, radius, thickness and chemical composition.
There is also the possibility to use other parameters to describe the
covering and filling factors to account for the 
inhomogeneities of the ejecta, but since we were interested only in the fluxes 
relative to \Hb, these parameters were not changed in the models. The code 
then integrates the emission over this shell giving the emergent 
emission-line spectrum.

The best fit parameters for CLOUDY are shown in Table 3 (temperature of the ionizing 
source and hydrogen density of the shell) and Table 4 
(chemical abundance of the shell). The intrinsic errors in the temperature and density
determinations are estimated to be 15 and 20\% respectively, but systematic 
errors are likely to dominate, due for example to the fact that we are 
assuming spherical symmetry, eventhough observations show this is not the case.
Figures 5 through 10 we 
present the comparison between the modeled synthetic spectra and the 
dereddened observations for days
85, 140, 172, 202, 250, 401, 506, 565, 639, 1381 and 1581 after the outburst.
The observations were dereddened with E(B-V)=0.32 
(\markcite{cho93}Chochol, D. et al. 1993).
Table 5 presents the dereddened observed fluxes of some selected 
lines compared to the model predictions. 

In order to directly compare the model and the data, the continuum was
subtracted in all the observed spectra (a technique particularly important for
the 1381 and 1581 days), and an arbitrary constant was added to set a constant 
continuum levels. 
We created synthetic spectra from the model output using a 
Gaussian function with a fixed FWHM of 30 \AA\ to approximate the
observed line widths.

In the figures comparing the models to the data, the
emission calculated from the photoionization models are
identified in the upper row. Only those lines with integrated
flux $\ge 10^{-3}$ relative to \Hb~ are shown. In those
cases where the lines are too close together only the brightest line from
the ``blend'' is labeled (although all the lines are marked).
The identification of the rest of the lines in a blend can be
found in the figure captions. From the figures it is clear that the brightest 
lines change from one day to another as the physical conditions of the nebula
evolve. The lower row labels mark the lines which are observed but 
not predicted by the models, this is either because they are not included 
in the code or because their integrated fluxes are lower than $10^{-3}$.  
This is the case, for instance, of the  [O~I] \lam\lam6300/63 lines, which 
are observed but for which the models predict even stronger lines of  
[S~III] \lam6312~and Fe X \lam6374.

\subsection{Abundances}

The abundances of the elements were initially taken from \markcite{aus96}A96,
and since no dust formation was reported in Nova Cygni 92 
(\markcite{Par95}Paresce
et al. 1995; \markcite{hay96}H96), we do not expect a significant depletion 
of heavy elements; we therefore assume that the mean chemical composition of 
the gas remains constant in time.

We found that the initial values given by A96 overestimated the strength 
of the C II \lam4267, N II \lam5680, [S~III] \lam6312, Ar III \lam7135,
plus some of the Fe and O lines.  The gas abundances (relative to Hydrogen, 
with H = 1 for the solar value) were then modified to 
improve the comparison with the observations as follows: 
S = 122.6 $\rightarrow$ S = 1,
N = 282 $\rightarrow$ N = 50, O = 110 $\rightarrow$ O = 80,
Ar = 58.2 $\rightarrow$ Ar = 5.0, and Fe = 16 $\rightarrow$ Fe = 8.
The changes in the Sulphur, Nitrogen and
Argon abundances are relatively large, but this is part due to the fact that
except for H, He, O, Ne, N and Fe, the
initial abundances from A96 were not properly constrained 
(\markcite{sta97}Starrfield 1997).

The Sulfur abundance was reduced to 1 (to be considered as an upper limit), 
in agreement with 
\markcite{mat93}Matheson et al. 1993, 
since the [S~III] \lam6312~ line should be negligible compared to 
[O~I] \lam6300. The Nitrogen abundance was reduced to 50 to fit the 
[N~II] \lam\lam5680/5755~ lines, which
provides a good match for the early days, but for day 250 and following days
the lines are slightly underestimated.  The abundance value of N is high when
compared to that found by \markcite{mat93}Matheson et al. (1993), 
but matches the value derived by \markcite{hay96}H96 based on the mid-infrared 
spectra of Nova Cygni 92. The Oxygen abundance was reduced from 110 to 80 to get 
a better fit for  [O~III] \lam\lam4959/5007.  We therefore increase the Ne/O 
ratio from 2.3 (\markcite{aus96}A96) to 3.1, in closer agreement with the 
value of Ne/O=4 found by \markcite{sal96}Salama et al. (1996).
The N/O ratio gives a value of 0.6, which is consistent with a white dwarf's
mass between 1.25 \msolar~and 1.35 \msolar~(\markcite{sta92}Starrfield et al. 
1992).
From the Ar~III \lam\lam7135/7007 lines we set an upper limit to the Ar 
abundance of 5.  Finally, the Fe abundance was set to fit the Fe~III \lam5270 and Fe~VI \lam5177 lines.

It is important to remember that the only abundances our models constrain are 
those for which we have emission line observations.
The abundances we modeled and used are, He=4.5, N=50.0,
O=80.0, Ne=250.0, S=1.0, Ar=5.0, Fe=8.0 (relative to their solar values in 
which hydrogen is unity).
They are within the $68\%$ confidence limits of those listed by A96,
except for N which is difficult to constrain since we do not have the strong
Nitrogen UV lines to compare (see \markcite{sho96}Shore et al. 1996). 
The abundances in Table 4 include the
values for other elements (following A96), that may not correspond to the 
best values. We do not expect, however, that changes in them will modify the
spectra. As \markcite{pol95}Politano et al. (1995) and \markcite{aus96}A96 
point out, the high abundances provide further support for 
the mixing scenario in  which core material from the ONeMg white dwarf 
is dredged up during the accretion process.
  
\subsection{Temperature and Luminosity of the White Dwarf}

The time evolution of the temperature of the ionizing source undergoes
three different phases (see Table 3): a rapid rise during days 140 through 
250 (at least),
reaching a plateau (in our data) at around day 400, and beginning to decline
after day 506. These phases are necessary to explain the different observed 
ionization stages as a function of time and they are consistent with the 
behavior of the X-ray flux observed with ROSAT by \markcite{kra96}Krautter et 
al. (1996). 

As it is explained by \markcite{kra96}Krautter et al. (1996) thermonuclear 
runaway models of the nova outburst (\markcite{sta89}Starrfield, 1989), 
predict that not all of the accreted material is ejected during the outburst.
The remaining material (ranging from 10\% to 90\%) would radiate at a 
constant luminosity. As mass is lost from this system and the white dwarf 
shrinks back to its equilibrium radius, the effective temperature of the 
remnant increases. \markcite{kra96}Krautter et al. (1996) related the decline
in the X-ray flux with a decrease of the white dwarf's temperature, due to 
the turnoff of thermonuclear burning at its surface.  
The temperature for the central source that we derived for the different 
models is consistent with the $90\%$ confidence temperature values obtained 
by A96. Our central source temperatures are lower in the rise phase than those
obtained by H96 based on mid-infrared data, but they are similar during the
plateau and decline phases.

In the models we used a simple approximation of a constant luminosity of 
L$_{\rm WD}$ = $10^{38}$ erg s$^{-1}$ for the central source to minimize the 
number of initial parameters to study. The constant bolometric 
luminosity phase depends on the mass of the white dwarf, and according to
\markcite{kra96}Krautter et al. (1996), lasts at least 511 days for Nova
Cygni 1992 (\markcite{sho96}see also Shore et al. 1996). A value of 
$10^{38}$ erg s$^{-1}$ was also used by H96 up to day 616, 
but it was reduced to $10^{35.8}$ erg s$^{-1}$ for day 849.
A luminosity of $10^{38}$ erg s$^{-1}$ is then clearly too high for our 
models corresponding to the 1381 and 1581 days. (See note on days 1381 and 
1581 on next section for further discussion).

\subsection{Radius, Density and Thickness of the Shell}

In all our models the physical radius of the shell (see Table 3) was derived 
from HST images (\markcite{par95}Paresce et al. 1995) taking the mean 
value between the major and minor axes of the observed elliptical ring and
adopting a distance of 2.8 kpc 
(\markcite{aus96}A96). For days 1381 and 1581 we assumed a mean expansion
velocity of 620 km s$^{-1}$.  This is smaller than the 1100 km s$^{-1}$
derived above from the line shape model, but it is not critical 
for the line ratios. 

As it is expected from a constant expansion velocity shell, the density 
decreases approximately as $t^{-2}$. Our results agree well with those found 
by A96, \markcite{hay96}Hayward et al. 1996 and \markcite{par95}Paresce et al.
1995.

The thickness ($\Delta$R) of the shell became an important parameter for 
days 1381 and 1581, when the shell changed from being radiation-bounded to 
matter-bounded. A similar behavior was found by \markcite{mor96}Morisset et 
al. (1996), where the shell became matter-bounded by day 800.  We chose a 
constant thickness of $10^{14.8}$ cm, which is the average between the lower 
and upper limit given by \markcite{aus96}A96. For days 1381 and 1581 this 
value is also consistent with the criteria followed by \markcite{hay96}H96 
who adopted a thickness equal to a value of a $10\%$ the radius of the shell.  
\markcite{sai96}Saizar et al. (1996) also suggest that since the [O~I] 
\lam\lam6300/63 lines trace the presence of neutral gas, the lines can be 
used to determine when the shell changes from ionization-bounded to 
matter-bounded.  The [O~I] lines are present in the spectra up to day 639, 
but they disappear by day 1381, indicating when the shell was 
matter-bounded.  

\subsection{Late Times}

The comparison between between observations and models for the first
seven epochs (i.~e. 85 through 565) is quite reasonable, and this can be
seen for instance in the [O~III] \lam\lam 5007+4959/\lam4363 line ratio (see
Figure~11). The ratio can be used to determine the electron temperature ($T_e$)
in the ejecta, for a range in the electron density 
of $\sim 10^3 - 7\times10^5$ \cc, and for values of $T_e$ of $\sim 10^3 -
10^5$ K (see e.~g. Osterbrock 1989). Departures from these ranges indicates
in general a more complex ionization structure. By day 565,
one begins to notice some difference between our simple models and the
observations. These differences are more noticeable for the last two epochs
of our observations, 1381 and 1581.

For days 1381 and 1581, we encountered several problems. First, the lack of consecutive 
observations during the 
1994$-$95 period makes it very difficult to follow the time evolution 
of the white dwarf temperature and the hydrogen density (the main parameters we
try to determine).  Second, as it was mentioned before, the luminosity of 
white dwarf is expected to decrease considerably after day 600.  H96 suggest
it goes down to $10^{35.8}$ erg s$^{-1}$ by day 849.  As we already mentioned
our approximation of a 
constant luminosity of $10^{38}$ erg s$^{-1}$ is clearly too high for those 
days. We ran several models with a lower value of $10^{35.8}$ erg s$^{-1}$ 
and the general tendency was that the [Ne~V]\lam3346 and [Ne~V]\lam3426
lines did not appear in the models, while they are clearly present in  
the observations.  Third, the ionizing spectrum at late times may depart
significantly from a blackbody. 
Fourth, our models for days 1381 and 1581 were very sensitive
to the thickness of the shell chosen, since the ionization structure of H, He, 
C, N and O depends strongly on the radius. Our simple approach of a single 
uniform density shell for these latter days, however, is not appropriate.
H96 in their modeling of the mid-infrared spectra, have used "multiple 
components" to reproduce the observations, i.e. three zones with different 
physical characteristics (dense, intermediate and diffuse). The contrast among
these three components becomes more drastic after approximately day 600.  
We have used a ``single'' component model up to day 639, with 
relatively good results. Nevertheless for days 1381 and 1581 this approximation
seems to break down.  For example, our synthetic spectra cannot reproduce
the [OIII](5007+4959)/4363 ratio, which is 
overestimated by a large factor. Since high ratios are indicative of low density 
material, the fact that the ratio is greatly overestimated indicate that
a single low-density component model is not valid at late times, when  
the emission is probably dominated by denser clumps.

\section{The Coronal Lines}

The spectra of Nova Cygni 1992 display some very high ionization lines of 
[Fe VI], [Fe VII] and [Fe X].
The highest ionization forbidden fine-structure emission lines or 
coronal lines are expected to arise in a high temperature low density gas;
whether this gas is heated by photoionization alone 
or by the interaction through shocks of the ejecta with the circumstellar
material, is still not known.  \markcite{aus96}A96 and \markcite{hay96}H96
found in their studies that photoionization alone fully accounts for the 
coronal lines seen in the optical and in the mid-infrared.  
A similar conclusion was reached by \markcite{wag96}Wagner and DePoy (1996) 
who concludes that the observed IR spectrum of Nova Cygni 1992 and the inferred 
electron temperature are inconsistent with an origin in a high-velocity 
shock-heated gas.  From HST observations \markcite{par95}Paresce et al. (1995)
also found that shocks are not responsible for a substantial part
of the emission.  On the other hand, \markcite{90}Greenhouse et al. 
(1990), determined from studies on QUVul that photoionization is insufficient 
to account for the spatial extent of the coronal-emitting region. 
Figure~12 shows the [Fe~VI] \lam\lam5146/77, [Fe~VII] \lam6087 + Ca~V \lam6087 
and [Fe~X] \lam6374 coronal emission lines as a function of time and they are 
compared with the behavior of the ``super-soft''X-Rays [0.1 $-$ 2.4 KeV] 
observed by ROSAT (\markcite{kra96}Krautter et al. 1996) and the values 
predicted by the models. On the one hand the similar behavior of the three 
quantities strongly suggests that an important fraction of the iron coronal 
lines could be due to photoionization.  On the other hand all the lines  are
found to be systematically underestimated which may indicate some contribution
from shocks.  It is also possible that the strength of the Fe lines reflects 
our chosen Fe abundance, which was reduced by a factor of two to match the 
spectra. Furthermore, the Fe X \lam6374 line is blended with the [O~I] \lam6363
line, so to obtain its flux we subtracted one third of the [O~I] \lam6300 line
from the blend (assuming too that the contribution from the [S~III] \lam6312 
line was negligible). It could be also the case that the coronal lines are 
produced  as a "skin effect" on the lower-density or rarified surfaces of 
denser clumps, which could explain the difference between the observed and 
modeled Fe lines. Multiple component models would be required to explore
this in more detail.

\section{Summary}

The spectroscopic time evolution of Nova Cygni 1992 has been studied
using 4 years of observations coupled with the synthetic 
spectral models, in order to learn more about the physical conditions
of its emitting gaseous shell. The line fluxes for the models were obtained
using a photoionization code (CLOUDY), where the chemical gas abundances
were kept {\it fixed} as a function of time.

The models are consistent with the results from similar studies
(\markcite{aus96}Austin et al. 1996), although they require less drastic
enhancements in the gas abundances of the heavy metals, in particular for
Sulphur, Argon (which were not modeled by A96) and Nitrogen.

The time evolution of the surface temperature of the ionizing
source (which it is assumed to be a black body) follows three phases (a rise,
a plateau and a decline) and it ranges from log(T) = 4.9 - 5.63, similar to 
what it is observed in the soft X-rays
(\markcite{kra96}Krautter et al. 1996).

The line profiles can be reproduced using a simple kinematical model, 
consisting 
of an equatorial ring and polar caps with slightly different densities 
(6.7$\times$10$^{7}$ - 9.5$\times$10$^{7}$ cm$^{-3}$ respectively), 
expanding at an average velocity of 1100~km/s.

The Fe Coronal emission lines follow also the time behavior
observed in the soft X-rays, which suggests a common energy source for their
photoionization. That the Fe line fluxes are slightly underestimated, may
indicate nevertheless that the contributions from shocks may also be important.

\vspace*{1.0cm}

\acknowledgements

AMM thanks MMO and IPAC for providing access to their facilities during the
completion of this work. We thank Gary Ferland for generously making 
available his photoionization code CLOUDY.
We thank A. Alpert for helping on the data reduction in the initial phase of 
the project and E. Friel for her useful comments.

\clearpage

\tablenum{1}
\pagestyle{empty}

\makeatletter
\def\jnl@aj{AJ}
\ifx\revtex@jnl\jnl@aj\let\tablebreak=\nl\fi
\makeatother

\begin{deluxetable}{lcccc}
\tablewidth{16.5cm}
\tablecaption{Log of Observations}
\tablehead{
\colhead{Date} &
\colhead{Day\tablenotemark{a}} & \colhead{Telescope} & 
\colhead{Wavelength (\AA)} &  
\colhead{Resolution (\AA)}}
\startdata
May{\hskip 5pt 15}, 92 & 85 & DAO 1.8m & 3500-8700 & 3.1 \\
Jul{\hskip 10pt 09}, 92 & 140 & DAO 1.8m & 3700-7500 & 2.2 \\
Jul{\hskip 10pt 12}, 92 & 143 & DAO 1.8m & 6500-6700 & 1.2 \\
Aug{\hskip 6pt 10}, 92 & 172 & DAO 1.8m & 3900-7300 & 3.1 \\
Sep{\hskip 8pt 09}, 92 & 202 & DAO 1.8m & 3300-6700 & 3.5 \\
Sep{\hskip 8pt 17}, 92 & 210 & DAO 1.8m & 4400-8000 & 6.5 \\
Oct{\hskip 8pt 27}, 92 & 250 & DAO 1.8m & 3300-6900 & 2.8 \\
Mar{\hskip 6pt 27}, 93 & 401 & DAO 1.8m & 4300-6900 & 2.2 \\
Jul{\hskip 10pt 10}, 93 & 506 & DAO 1.8m & 3300-6400 & 6.1 \\
Sep{\hskip 8pt 07}, 93 & 565 & DAO 1.8m & 3300-7300 & 5.3 \\
Nov{\hskip 7pt 20}, 93 & 639 & KPNO 2.1m & 4400-8000 & 5.8 \\
Dec{\hskip 8pt 02}, 95 & 1381 & MMT & 3200-8900 & 5.8 \\
Dec{\hskip 8pt 03}, 95 & 1382 & MMT & 4000-5500 & 2.0 \\
Jun{\hskip 9pt 19}, 96 & 1581 & MMT & 3300-8900 & 5.8 \\
\tablenotetext{a}{Days after outburst}
\enddata
\end{deluxetable}

\tablenum{2}
\pagestyle{empty}

\makeatletter
\def\jnl@aj{AJ}
\ifx\revtex@jnl\jnl@aj\let\tablebreak=\nl\fi
\makeatother
\begin{deluxetable}{lr}
\tablewidth{8.0cm}
\tablecaption{Model Line Profile}
\tablehead{
\colhead{Parameter} & \colhead{Value}}
\startdata
Angle of View\tablenotemark{a} & 33.0 \nl
v$_{\rm max}$\tablenotemark{b} & 1100.0 \nl
v$_{\rm min}$\tablenotemark{b} & 1000.0 \nl
Ring angle\tablenotemark{c} & 6 \nl
Ring enhancement\tablenotemark{d} & 2.0 \nl
Polar Cap angle\tablenotemark {e} & 10 \nl
Polar Cap enhancement\tablenotemark{d} & 2.9 \nl
\tablenotetext{~}{distance = 2.8 Kpc (image)}
\tablenotetext{~}{time after explosion=1.5$\times$10$^{7}$ s (image)}
\tablenotetext{a}{from the pole ($^\circ$)} 
\tablenotetext{b}{km~s$^{-1}$}
\tablenotetext{c}{angle above/below the equator ($^\circ$)} 
\tablenotetext{d}{emission enhancement over shell}
\tablenotetext{e}{angle from the pole($^\circ$)}
\enddata
\end{deluxetable}

\tablenum{3}
\pagestyle{empty}

\makeatletter
\def\jnl@aj{AJ}
\ifx\revtex@jnl\jnl@aj\let\tablebreak=\nl\fi
\makeatother

\begin{deluxetable}{lrrrr}
\tablewidth{12.0cm}
\tablecaption{Model Parameters\tablenotemark{a}}
\tablehead{
\colhead{Date} &
\colhead{Day\tablenotemark{b}} & \colhead{log(T)\tablenotemark{c}} & 
\colhead{log(N$_{\rm H}$)\tablenotemark{d}} &  
\colhead{log(R$_{\rm in}$)\tablenotemark{e}}}
\startdata
May{\hskip 5pt 15}, 92 & 85 & 4.95 & 7.90 & 14.65 \\
Jul{\hskip 10pt 09}, 92 & 140 & 5.02 & 7.80 & 14.87 \\
Aug{\hskip 6pt 10}, 92 & 172 & 5.10 & 7.75 & 14.95 \\
Sep{\hskip 8pt 09}, 92 & 202 & 5.36 & 7.64 & 15.02 \\
Oct{\hskip 8pt 27}, 92 & 250 & 5.52 & 7.46 & 15.13 \\
Mar{\hskip 6pt 27}, 93 & 401 & 5.62 & 7.20 & 15.34 \\
Jul{\hskip 10pt 10}, 93 & 506 & 5.63 & 6.80 & 15.43 \\
Sep{\hskip 8pt 07}, 93 & 565 & 5.54 & 6.60 & 15.48 \\
Nov{\hskip 7pt 20}, 93 & 639 & 5.48 & 6.40 & 15.53 \\
Dec{\hskip 8pt 02}, 95 & 1381 & 4.88 & 4.46 & 15.86 \\
Jun{\hskip 9pt 19}, 96 & 1581 & 4.90 & 4.40 & 15.92 \\
\tablenotetext{a}{L$_{\rm WD}$ = 10$^{38}$ ergs s$^{-1}$, 
$\Delta$R = 10$^{14.8}$ cm, $^{\rm b}$ Days after outburst, \\
$^{\rm c}$ Black Body Temperature (K), $^{\rm d}$ Hydrogen Density (cm$^{-3}$),
$^{\rm e}$ Inner Radius (cm)}
\enddata
\end{deluxetable}

\tablenum{4}
\pagestyle{empty}

\makeatletter
\def\jnl@aj{AJ}
\ifx\revtex@jnl\jnl@aj\let\tablebreak=\nl\fi
\makeatother
\begin{deluxetable}{lr}
\tablewidth{15pc}
\tablecaption{Abundances\tablenotemark{a}} 
\tablehead{
\colhead{Element} & \colhead{Abundance}}
\startdata
He & 4.5 \nl
C & 70.6 \nl
N & 50.0 \nl
O & 80.0 \nl
Ne & 250.0 \nl
Na & 37.4 \nl
Mg & 129.4 \nl
Al & 127.5 \nl
Si & 146.6 \nl
S & 1.0 \nl
Ar & 5.0 \nl
Ca & 46.8 \nl
Fe & 8.0 \nl
Ni & 36.0 \nl
\enddata
\tablenotetext{a}{Relative to H$_{\odot}$ = 1}
\end{deluxetable}

\tablenum{5}
\pagestyle{empty}

\makeatletter
\def\jnl@aj{AJ}
\ifx\revtex@jnl\jnl@aj\let\tablebreak=\nl\fi
\makeatother
\begin{deluxetable}{lllrrrrrr}
\tablewidth{15.0cm}
\tablecaption{Dereddened\tablenotemark{a}~ Observed Fluxes compared to Model
\tablenotemark{b}}
\tablehead{
\colhead{Ion} & \colhead{$\lambda$ (\AA)} & 
\colhead{} &  
\colhead{85\tablenotemark{c}} &
\colhead{140\tablenotemark{c}} & 
\colhead{172\tablenotemark{c}} & 
\colhead{202\tablenotemark{c}} &
\colhead{250\tablenotemark{c}} & 
\colhead{401\tablenotemark{c}}}
\startdata
[Ne~V] & 3345.9\tablenotemark{d} & F$_{\rm ob}$ & \nodata & \nodata & \nodata 
& 3.3 & 9.5 & \nodata \nl
 & & F$_{\rm mod}$ & 0.005 & 0.02 & 0.1 & 3.0 & 9.5 & 20.6 \nl
[Ne~V] & 3425.8 & F$_{\rm ob}$ & \nodata & \nodata & \nodata & 7.7 & 23.9 & 
\nodata \nl
 & & F$_{\rm mod}$ & 0.01 & 0.06 & 0.3 & 8.0 & 25.7 & 55.7 \nl
[Ne~III] & 3868.7\tablenotemark{e} & F$_{\rm ob}$ & 3.4 & 4.4 & \nodata & 5.8 & 1
0.1 & \nodata \nl
 & & F$_{\rm mod}$ & 3.5 & 4.7 & 6.2 & 9.2 & 10.4 & 17.0 \nl
[Ne~III] & 3967.5\tablenotemark{f} & F$_{\rm ob}$ & 1.2 & 1.5 & 1.9 & 1.9 & 3.1 &
 \nodata \nl
 & & F$_{\rm mod}$ & 1.1 & 1.4 & 1.9 & 2.8 & 3.2 & 5.2 \nl
H$\beta$ & 4861.3 & F$_{\rm ob}$ & 1.0 & 1.0 & 1.0 & 1.0 & 1.0 & 1.0 \nl
 & & F$_{\rm mod}$ & 1.0 & 1.0 & 1.0 & 1.0 & 1.0 & 1.0 \nl
[O~III] & 4958.9\tablenotemark{g} & F$_{\rm ob}$ & 0.6 & 0.7 & 0.8 & 1.0 & 1.6 & 
3.0 \nl
 & & F$_{\rm mod}$ & 0.6 & 0.7 & 0.9 & 1.3 & 1.6 & 2.8 \nl
[O~III] & 5006.9\tablenotemark{g} & F$_{\rm ob}$ & 2.1 & 2.2 & 2.6 & 3.2 & 5.0 & 
8.9 \nl
 & & F$_{\rm mod}$ & 1.7 & 2.2 & 2.8 & 4.1 & 4.7 & 8.3 \nl
[N~II] & 5680\tablenotemark{h} & F$_{\rm ob}$ & 0.08 & 0.06 & 0.05 & 0.07 & 
0.1 & 0.1 \nl
 & & F$_{\rm mod}$ & 0.06 & 0.05 & 0.05 & 0.04 & 0.03 & 0.03 \nl
[N~II] & 5755\tablenotemark{h} & F$_{\rm ob}$ & 0.1 & \nodata & 0.1 & 0.1 & 0.2 & \nodata \nl
 & & F$_{\rm mod}$ & 0.1 & 0.1 & 0.08 & 0.05 & 0.06 & 0.09 \nl
H$\alpha$ & 6562.8\tablenotemark{i} & F$_{\rm ob}$ & 4.5 & 4.2 & 3.8 & 3.3 & 2.7 
& 2.6 \nl
 & & F$_{\rm mod}$ & 4.8 & 3.7 & 3.3 & 2.9 & 2.8 & 2.7 \nl
\tablenotetext{a}{E(B-V)=0.32. $^{\rm b}$ Fluxes of the selected lines used to co
nstrain the parameters. 
$^{\rm c}$ Days after outburst. 
Blended lines are the following: $^{\rm d}$ [Ne~III] 3343, $^{\rm e}$ H8 3889,
$^{\rm f}$ H$\epsilon$ 3970, 
$^{\rm g}$ with each other,
$^{\rm h}$ with each other and with [Fe~VII] 5721,
$^{\rm i}$ [N~II] 6548/83} 
\enddata
\end{deluxetable}

\tablenum{5 (Cont.)}
\pagestyle{empty}

\makeatletter
\def\jnl@aj{AJ}
\ifx\revtex@jnl\jnl@aj\let\tablebreak=\nl\fi
\makeatother
\begin{deluxetable}{lllrrrrr}
\tablewidth{15.0cm}
\tablecaption{Dereddened\tablenotemark{a}~ Observed Fluxes compared to Model
\tablenotemark{b}}
\tablehead{
\colhead{Ion} & \colhead{$\lambda$ (\AA)} & 
\colhead{} &  
\colhead{506\tablenotemark{c}} &
\colhead{565\tablenotemark{c}} & 
\colhead{639\tablenotemark{c}} & 
\colhead{1381\tablenotemark{c}} &
\colhead{1581\tablenotemark{c}}}  
\startdata
[Ne~V] & 3345.9\tablenotemark{d} & F$_{\rm ob}$ & 31.4 & 21.9 & \nodata & 0.5 & 0
.4 \nl
 & & F$_{\rm mod}$ & 32.3 & 24.3 & 18.2 & 1.3 & 0.9 \nl
[Ne~V] & 3425.8 & F$_{\rm ob}$ & 87.5 & 67.1 & \nodata & 1.7 & 1.4 \nl
 & & F$_{\rm mod}$ & 87.5 & 65.8 & 49.4 & 3.5 & 2.6 \nl
[Ne~III] & 3868.7\tablenotemark{e} & F$_{\rm ob}$ & 12.0 & 9.9 & \nodata & 1.0 & 
0.9 \nl
 & & F$_{\rm mod}$ & 17.4 & 15.5 & 14.2 & 3.6 & 1.8 \nl
[Ne~III] & 3967.5\tablenotemark{f} & F$_{\rm ob}$ & 3.8 & 3.1 & \nodata & 0.5 & 0
.6 \nl
 & & F$_{\rm mod}$ & 5.3 & 4.8 & 4.4 & 1.1 & 0.5 \nl
H$\beta$ & 4861.3 & F$_{\rm ob}$ & 1.0 & 1.0 & 1.0 & 1.0 & 1.0 \nl
 & & F$_{\rm mod}$ & 1.0 & 1.0 & 1.0 & 1.0 & 1.0 \nl
[O~III] & 4958.9\tablenotemark{g} & F$_{\rm ob}$ & 4.2 & 4.9 & 8.1 & 1.2 & 0.9
 \nl
 & & F$_{\rm mod}$ & 3.7 & 4.1 & 4.8 & 1.3 & 0.7 \nl
[O~III] & 5006.9\tablenotemark{g} & F$_{\rm ob}$ & 12.1 & 13.8 & 24.8 & 3.7 
& 2.3 \nl
 & & F$_{\rm mod}$ & 11.2 & 12.4 & 14.4 & 4.0 & 2.0 \nl
[N~II] & 5680 & F$_{\rm ob}$ & 0.1 & 0.1 & 0.07 & \nodata & \nodata \nl
 & & F$_{\rm mod}$ & 0.03 & 0.04 & 0.04 & \nodata & \nodata \nl
[N~II] & 5755 & F$_{\rm ob}$ & 0.3 & 0.3 & 0.2 & \nodata & \nodata \nl
 & & F$_{\rm mod}$ & 0.10 & 0.09 & 0.08 & \nodata & \nodata \nl
H$\alpha$ & 6562.8\tablenotemark{h} & F$_{\rm ob}$ & \nodata & 2.8 & 3.8 & 
2.6 & 3.3 \nl
 & & F$_{\rm mod}$ & 2.7 & 2.7 & 2.7 & 2.6 & 2.6 \nl
\tablenotetext{a}{E(B-V)=0.32. $^{\rm b}$ Fluxes of the selected lines used to co
nstrain the parameters. 
$^{\rm c}$ Days after outburst. 
Blended lines are the following: $^{\rm d}$ [Ne~III] 3343, $^{\rm e}$ H8 3889,
$^{\rm f}$ H$\epsilon$ 3970, \\
$^{\rm g}$ with each other, 
$^{\rm h}$ with each other and with [Fe~VII] 5721,
$^{\rm i}$ [N~II] 6548/83}
\enddata
\end{deluxetable}

\clearpage

\figcaption [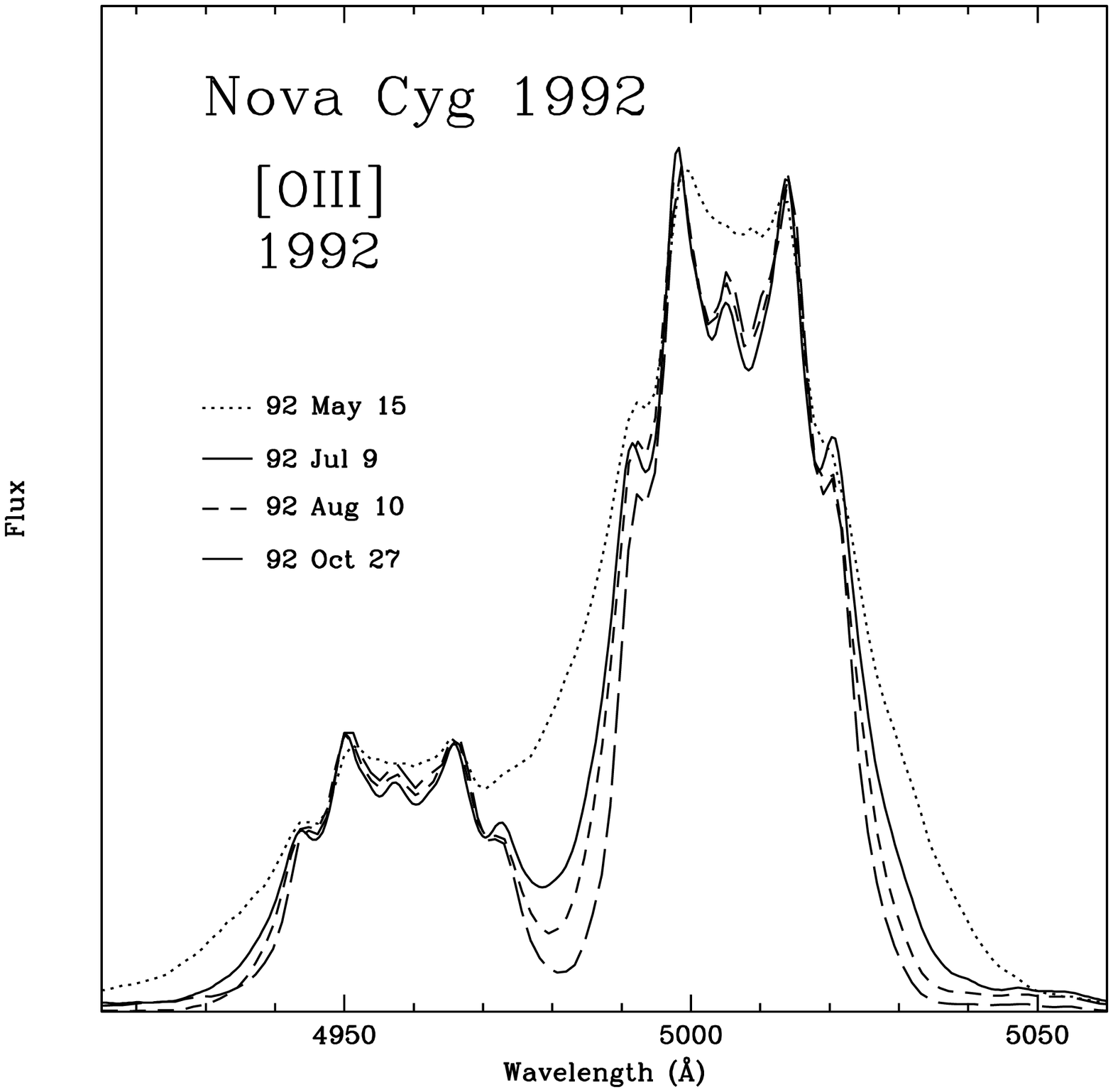]{From the 1992 spectra, the [O~III] 
\lam\lam4959/5007 emission of Nova Cygni 1992. Note the steady decline 
of width.}
%\label{fig1}}

\figcaption [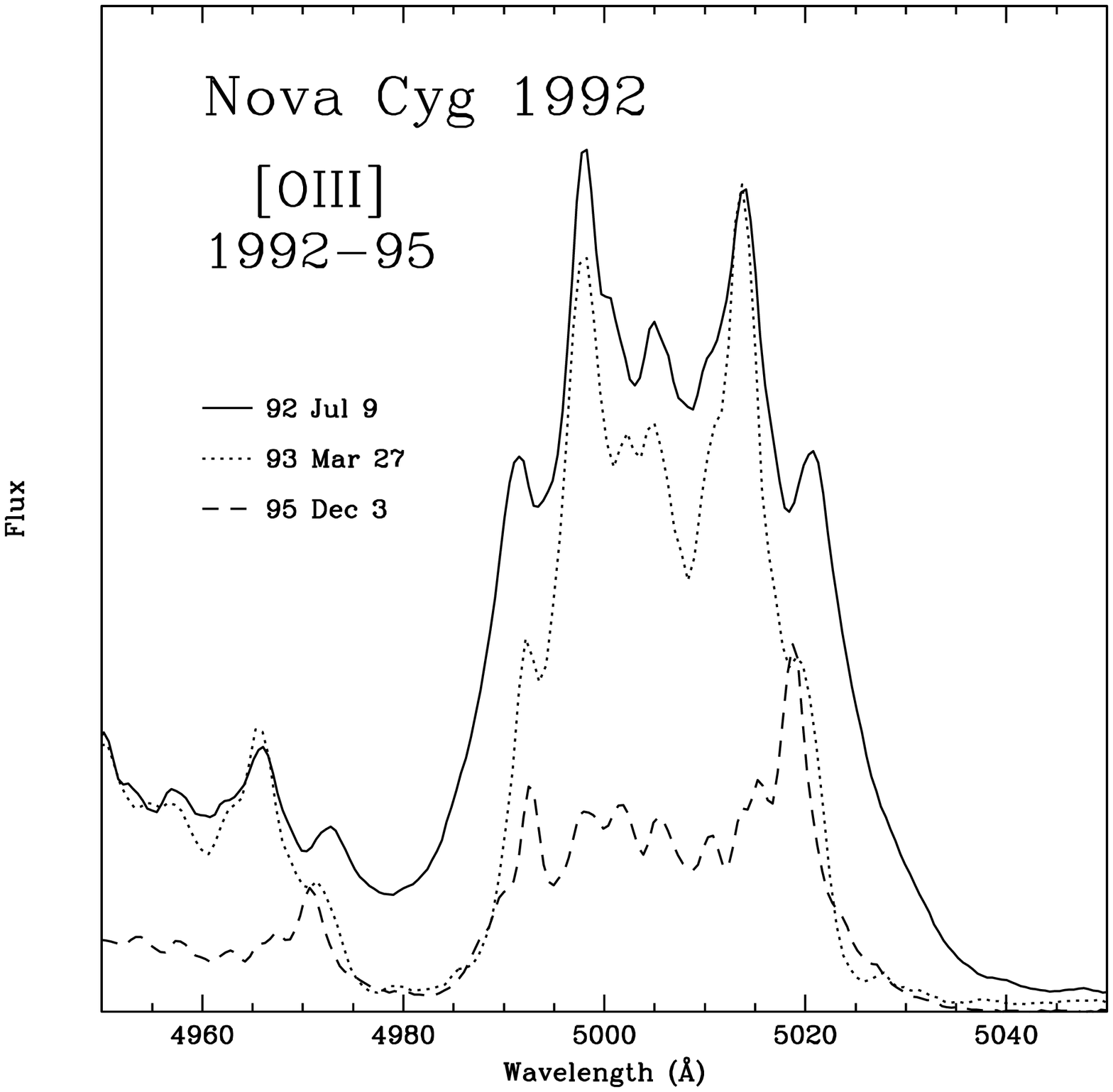]{The [O~III] \lam\lam4959/5007 emission of 
Nova Cygni 1992 from 1992 to 1995.}
%\label{fig2}}

\figcaption [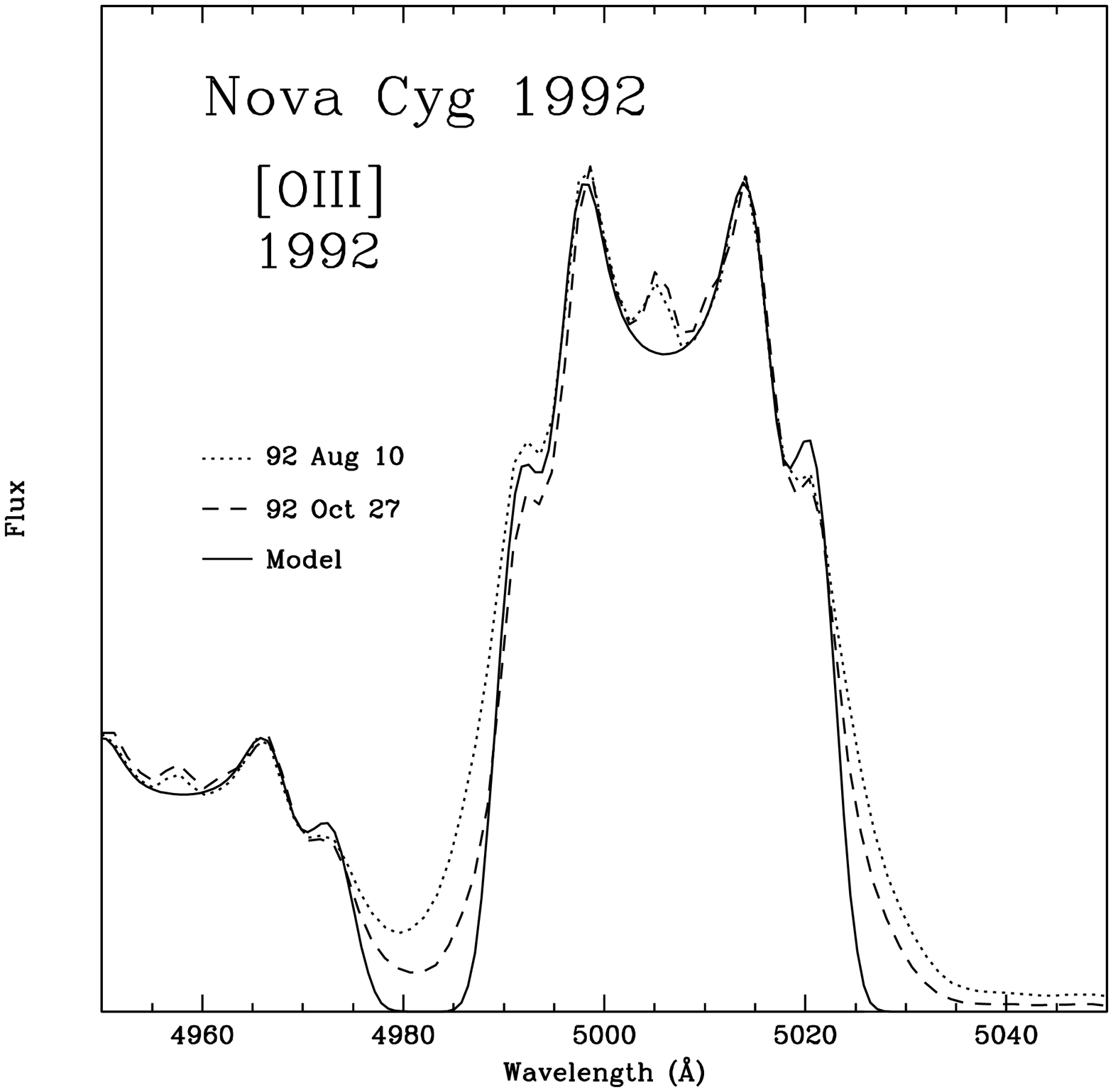]{The characteristic line profile of 
the \lam5007 emission line of Nova Cygni 1992 for two different days is 
compared with a simple kinematical model 
of an expanding ring and polar caps.}
%\label{fig3}}

\figcaption[moro_martin_fig04ab.ps]{The simple kinematical model
of an expanding ring with polar caps as projected onto the sky at the [O~III]  
\lam\lam5007 emission (left), compared with an archive 1994 HST FOC observation
of Cygni 1992 at the same wavelength range (right). 
The image model exaggerates perhaps the polar
caps but overall it resembles the observations.}
%\label{fig4}}

\figcaption [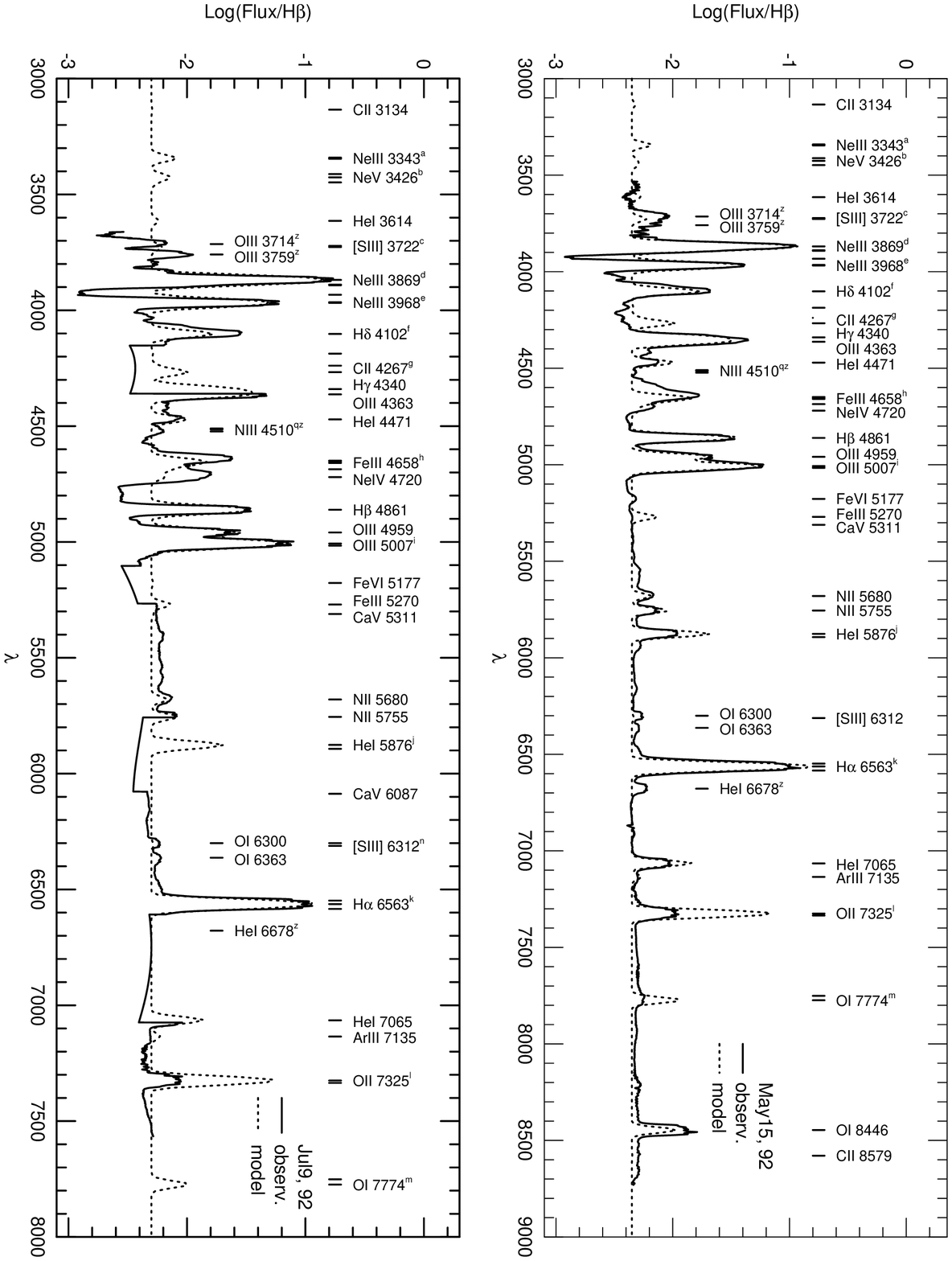]{A comparison between the observed (solid 
line) and synthetic (broken line)
spectra of Nova Cygni 1992 for four different days.  
Observations were corrected for reddening with E(B-V)=0.32
The vertical bars mark all the predicted lines $\ge 10^{-3}$
of H$\beta$. The top row labels correspond to the models,
which are essentially the same as the observations. The exceptions, i.e. 
which are too weak or not in the models, are
marked by the lower labels, as is the case for [O~I] \lam\lam6300/63 lines,
which it is overwhelmed in the models
by stronger [S~II] \lam6312 and Fe~X \lam6374 lines.
(top) May 15, 92. The superscripts indicate blended lines: 
%%$^{\rm a}$ OIII 3341 \& NeV 3346, 
{\it a}: OIII 3341 \& NeV 3346, 
%%$^{\rm b}$ OIV 3412 \& HeI 3448,
{\it b}: OIV 3412 \& HeI 3448,
%%$^{\rm c}$ OII 3727, 
{\it c}: OII 3727,
%%$^{\rm d}$ HeI 3889 \& FeV 3892, 
{\it d}: HeI 3889 \& FeV 3892,
%%$^{\rm e}$ CaII 3933 \& HeI 3965, 
{\it e}: CaII 3933 \& HeI 3965,
%%$^{\rm f}$ CIII 4187, 
{\it f}: CIII 4187,
%%$^{\rm g}$ NII 4239, 
{\it g}: NII 4239,
%%$^{\rm h}$ CIII 4649, OII 4651, CIV 4659, HeII 4686, \& NeIV 4720,
{\it h}: CIII 4649, OII 4651, CIV 4659, HeII 4686, \& NeIV 4720,
%%$^{\rm i}$ HeI 5016, 
{\it i}: HeI 5016, 
%%$^{\rm j}$ NaD 5893, 
{\it j}: NaD 5893, 
%%$^{\rm k}$ NII 6548 \& NII 6584, 
{\it k}: NII 6548 \& NII 6584, 
%%$^{\rm l}$ [ArIV] 7335, 
{\it l}: [ArIV] 7335,
%%$^{\rm m}$ ArIII 7751, 
{\it m}: ArIII 7751,
%%$^{\rm *}$ not in the models.
{\it q}: NIII 4514/23 and 
{\it z}: not in the models.
(bottom) July 9, 92.  Same blends as in Figure 5 (top) with 
{\it n}: OI 6300.}
%\label{fig5}}

\figcaption [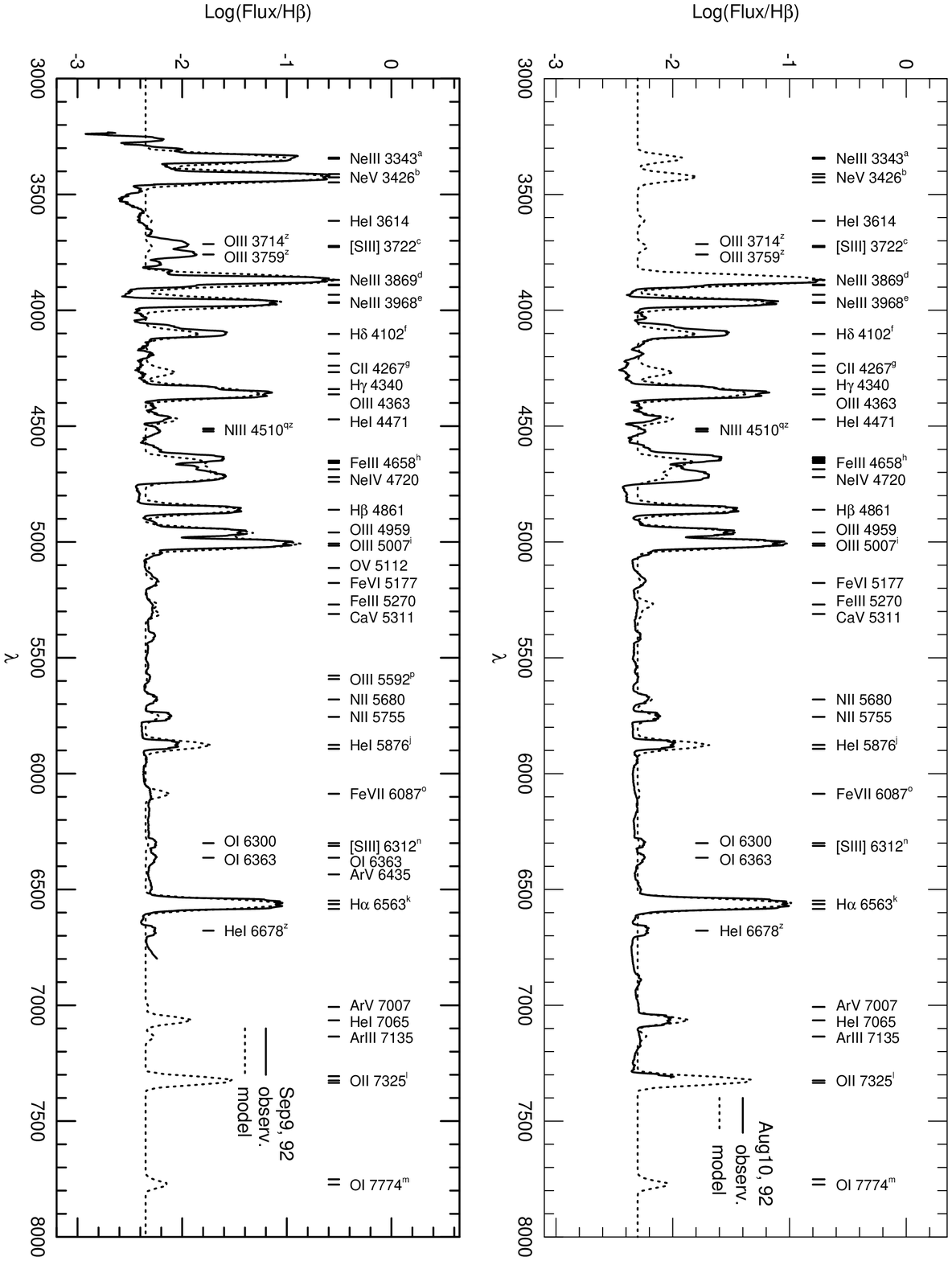]
{
(top) August 10, 92.  Same blends as in Figure 5 (top) with 
{\it n}: OI 6300 and {\it o}: CaV 6087.
(bottom) September 9, 92.  Same blends as in Figure 5 (top) with 
{\it h}:
including also ArIV 4740, 
{\it l}:
including also CaII 7306,
{\it n}: OI 6300, {\it o}: CaV 6087 and 
{\it p}: OI 5577.}
%\label{fig6}}

\figcaption [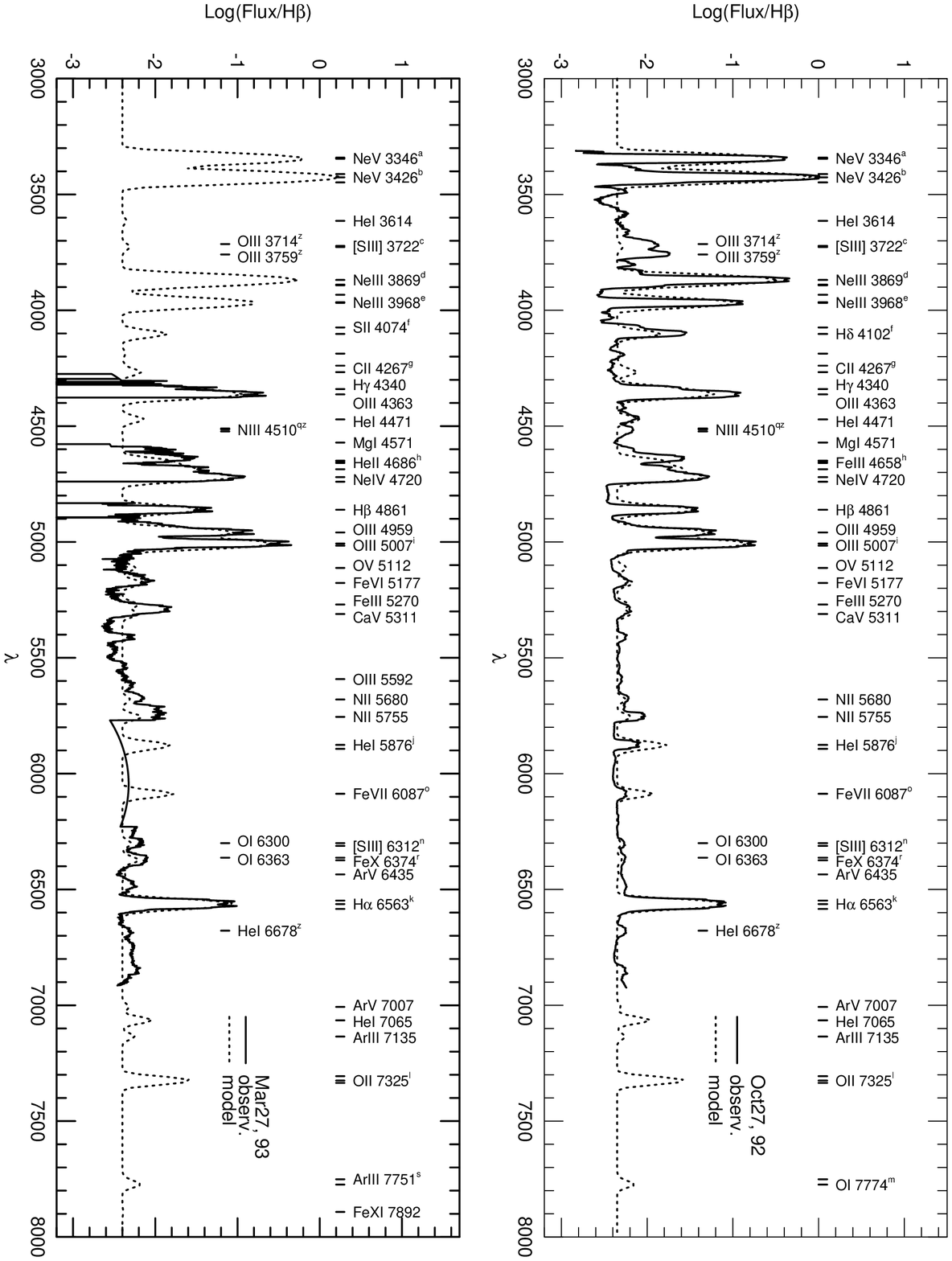]
{(top) October 27, 92.  Same blends as in Figure 5 (top) with 
{\it f}:
including also SII 4074,
{\it h}:
including also ArIV 4740, 
{\it l}:
including also CaII 7306,
{\it n}: OI 6300, {\it o}: CaV 6087, 
{\it p}: OI 5577, and 
{\it r}: OI 6363.
(bottom) March 27, 93.  Same blends as in Figure 5 (top) with 
{\it h}: 
including also FeIII 4658 and 
{\it s}: OI 7774.}
%\label{fig7}}

\figcaption [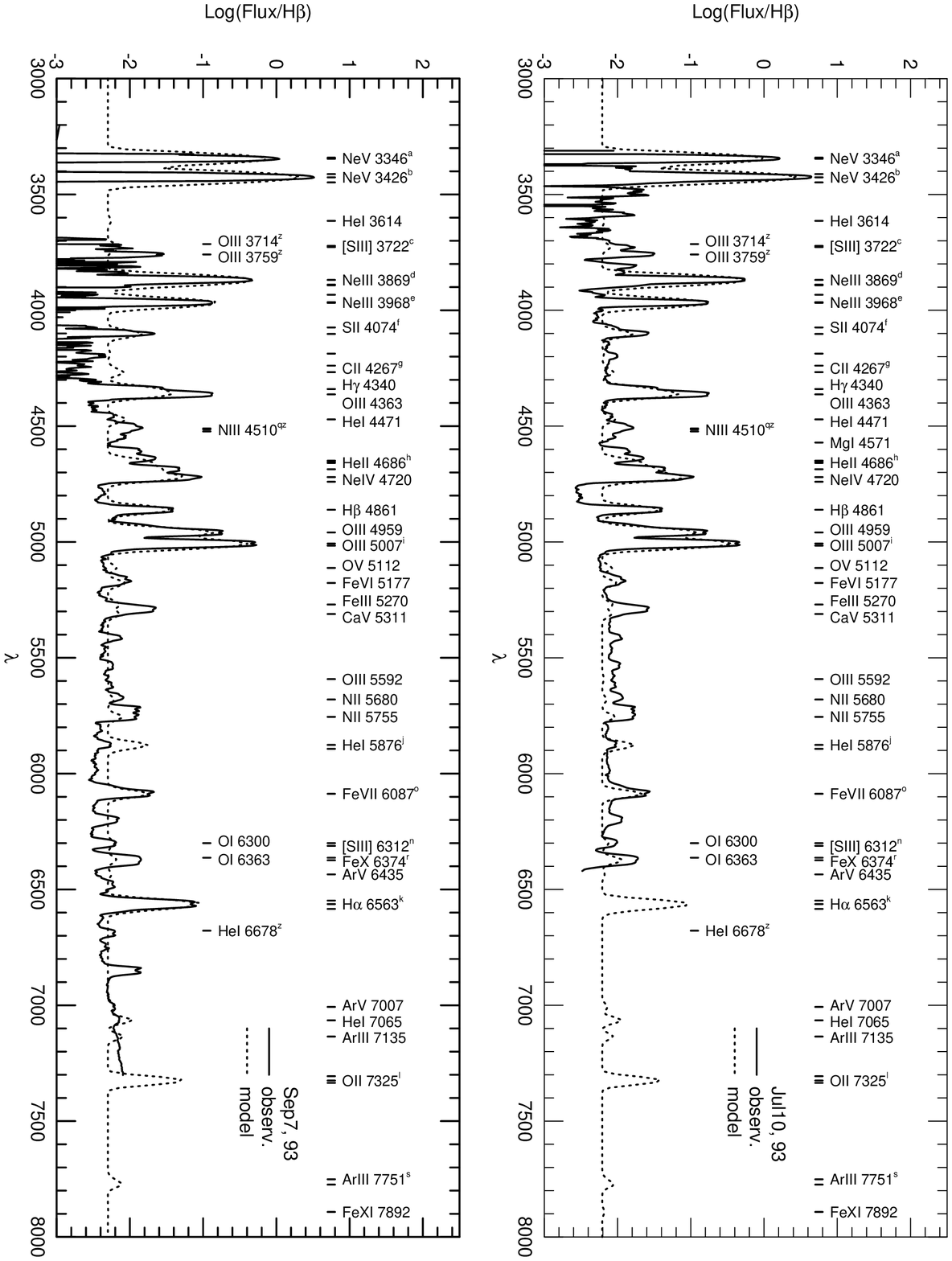]
{(top) July 10, 93.  Same blends as in Figure 5 (top) with {\it n}: OI 6300.
(bottom) September 7, 93.  Same blends as in Figure 5 (top) with {\it n}: 
OI 6300.}
%\label{fig8}}

\figcaption [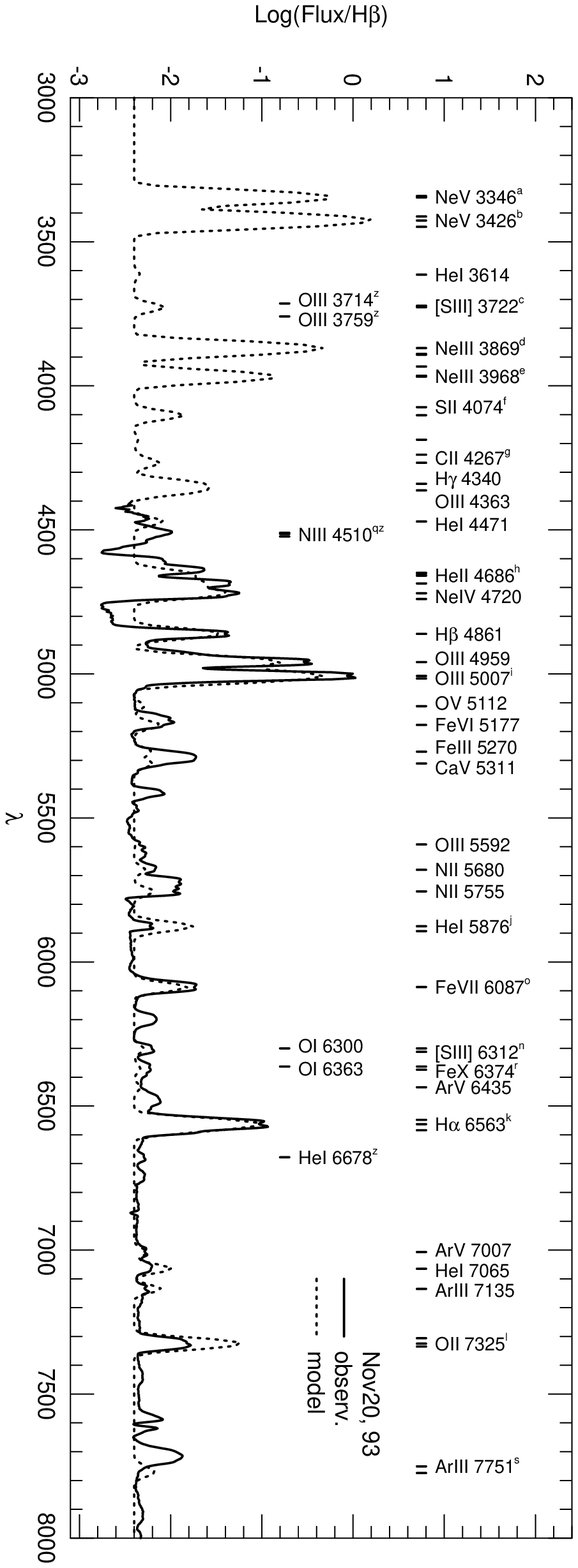]
{November 20, 93.  Same blends as in Figure 5 (top) with {\it n}: OI 6300.}    
%\label{fig9}}

\figcaption [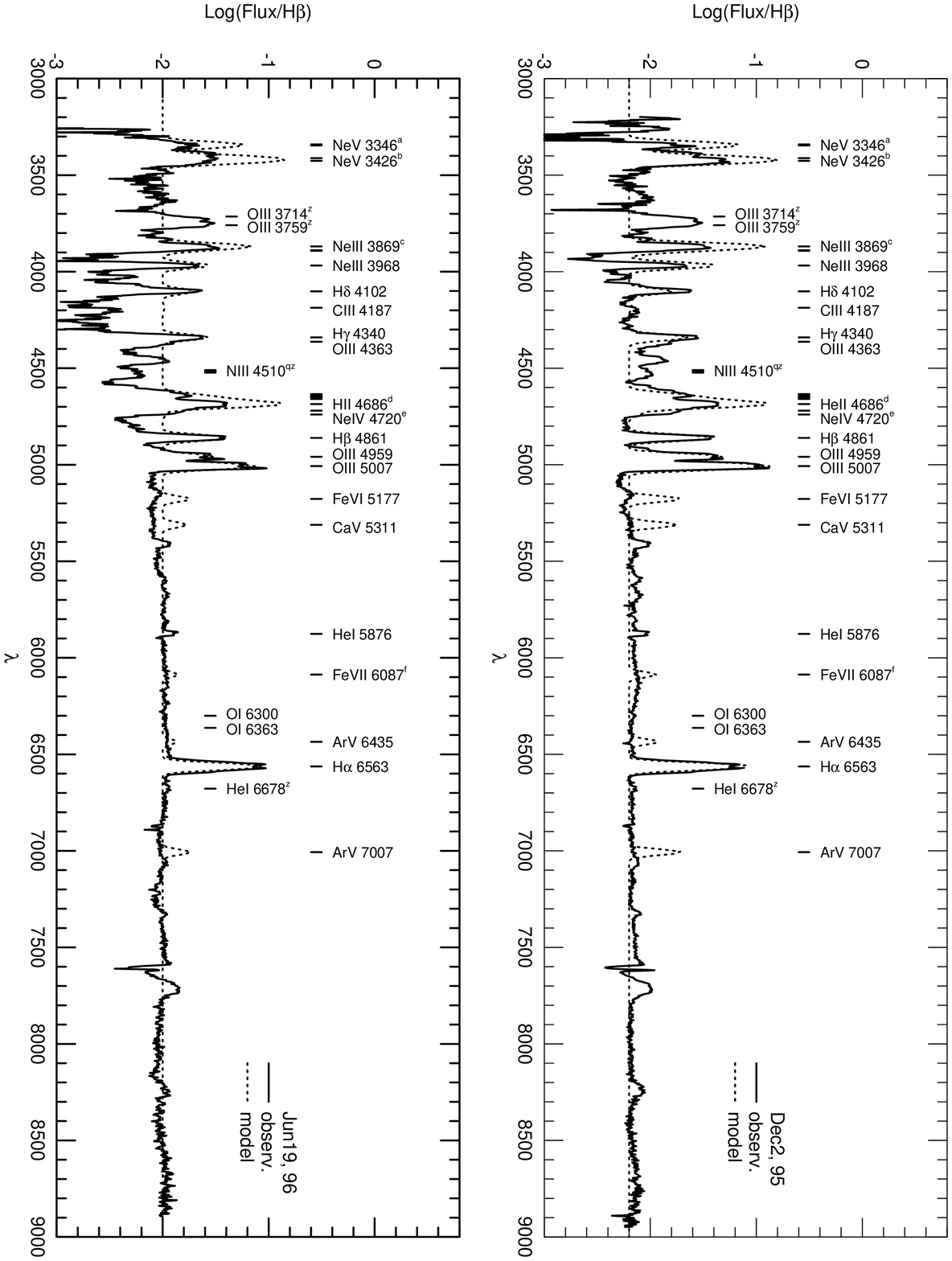]
{(top) December 2, 95.  Blended as follows: 
%%$^{\rm a}$ OIII 3341, 
{\it a}: OIII 3341,
%%$^{\rm b}$ OIV 3412, 
{\it b}: OIV 3412, 
%%$^{\rm c}$ HeI 3889 \& FeV 3892, 
{\it c}: HeI 3889 \& FeV 3892,
%%$^{\rm d}$ NIII 4640/34/42, CIII 4649, OII 4651 \& CIV 4659, 
{\it d}: NIII 4640/34/42, CIII 4649, OII 4651 \& CIV 4659,
%%$^{\rm e}$ ArIV 4740, 
%%$^{\rm f}$ CaV 6087.
{\it e}: ArIV 4740, 
{\it f}: CaV 6087 and
{\it q}: NIII 4514/23.
(bottom) June 19, 96.  Same blends as in Figure 5 (top) with {\it h}: 
including also FeIII 4658 and 
{\it s}: OI 7774}
%\label{fig10}}

\figcaption [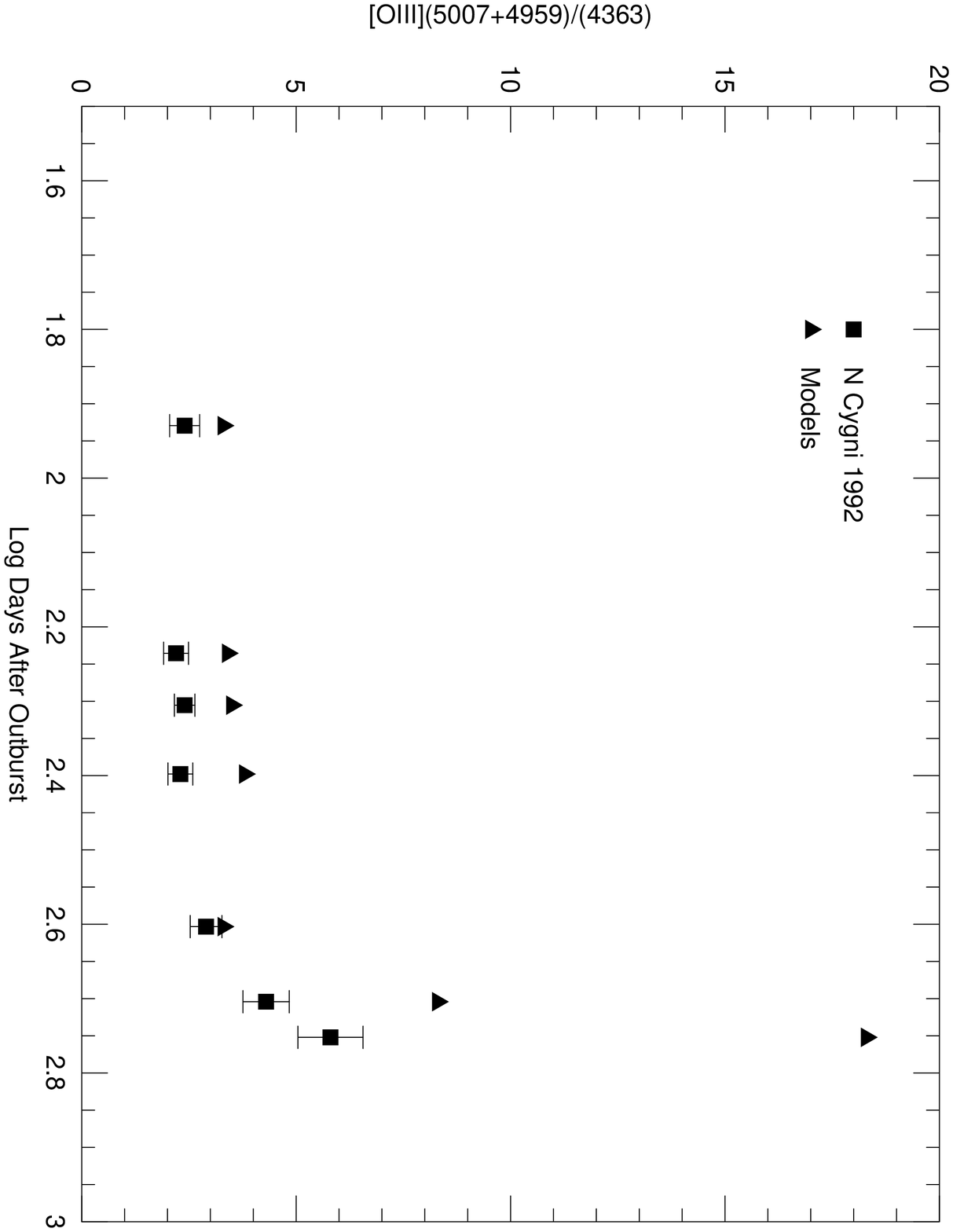]{Comparison between the observations 
and nebular models
of the [O~III] \lam\lam 5007+4959/\lam 4363 line ratio for our first
eight days (i.~e. 85 through 565). The models follow a similar trend early on
but they begin to depart at day 565, indicating perhaps that the ionization
structure of the ejecta becomes more complex.}
%\label{fig11}}

\figcaption [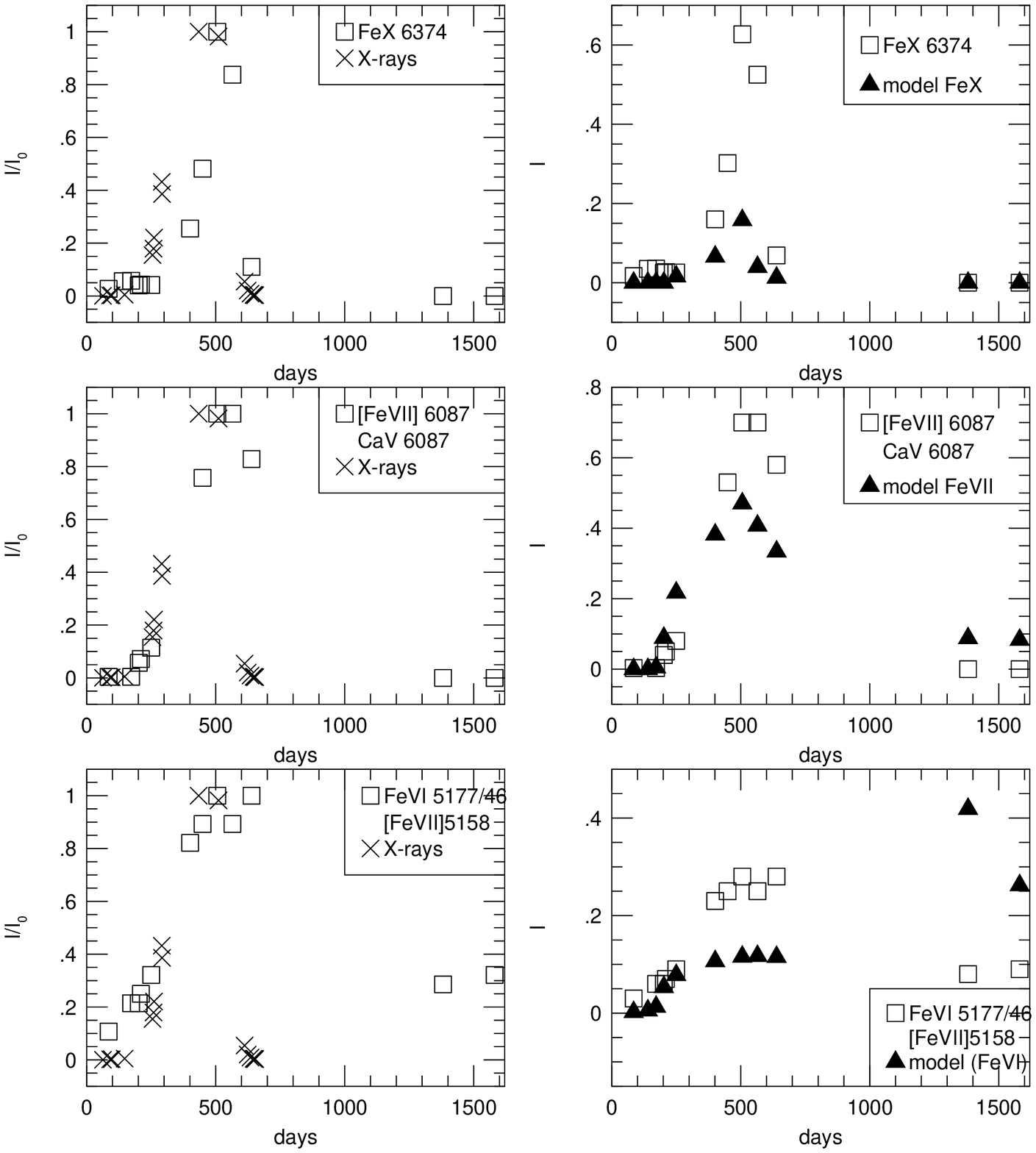]
{The time evolution of the Fe Coronal emission lines.  In the figures on the
left, the optical observations (open squares) are compared with the soft
X-rays (crosses).  The y-axis represents the intensity of the optical and
X-rays lines, each normalized to its maximum.  In the figures on the right,
the optical observations are compared with the Fe emission lines prediced by
the photoionization models (solid triangles). In this case, the y-axis
represents the intensity normalized to \Hb.}
%\label{fig12}}

\clearpage

\begin{figure}[ht]
\plotone{moro_martin_fig01.ps}
\end{figure}

\begin{figure}[ht]
\plotone{moro_martin_fig02.ps}
\end{figure}
 
\begin{figure}[ht]
\plotone{moro_martin_fig03.ps}
\end{figure}

\begin{figure}[ht]
\plottwo{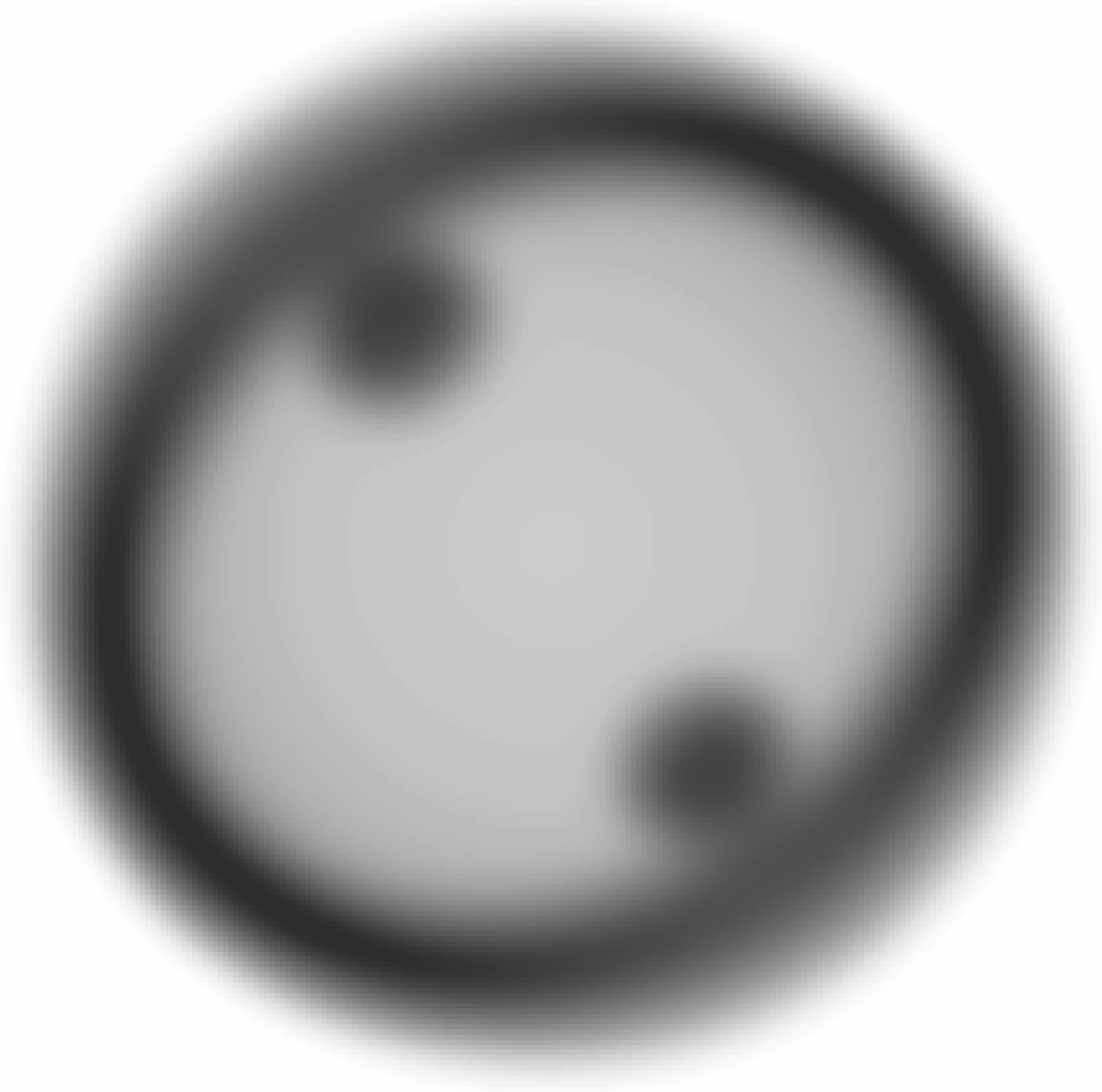}{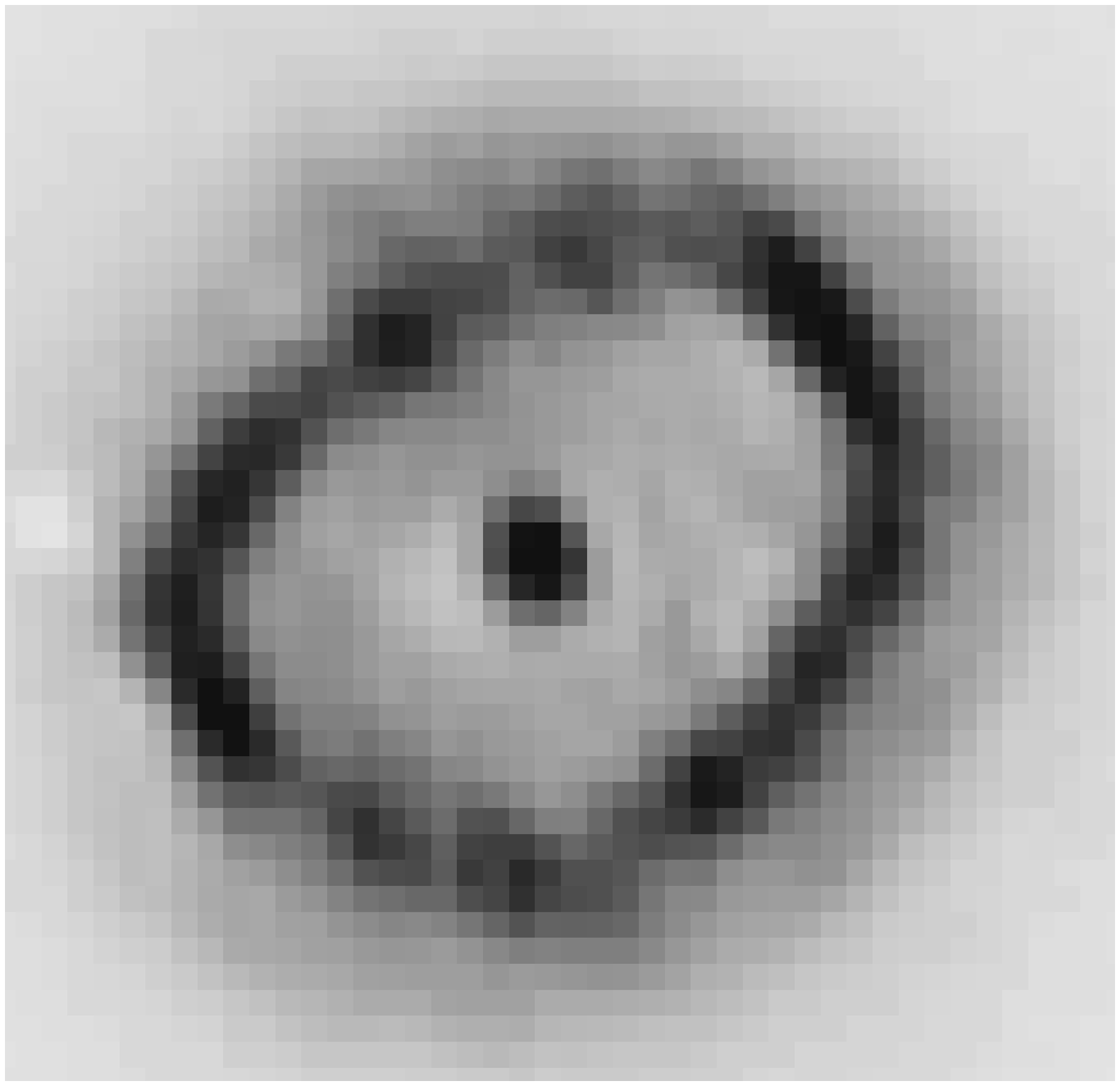}
\end{figure}

\begin{figure}[ht]
\plotone{moro_martin_fig05.ps}
\end{figure}

\begin{figure}[ht]
\plotone{moro_martin_fig06.ps}
\end{figure}

\begin{figure}[ht]
\plotone{moro_martin_fig07.ps}
\end{figure}

\begin{figure}[ht]
\plotone{moro_martin_fig08.ps}
\end{figure}

\begin{figure}[ht]
\plotone{moro_martin_fig09.ps}
\end{figure}

\begin{figure}[ht]
\plotone{moro_martin_fig10.ps}
\end{figure}

\begin{figure}[ht]
\plotone{moro_martin_fig11.ps}
\end{figure}

\begin{figure}[ht]
\plotone{moro_martin_fig12.ps}
\end{figure}

\end{document}